\documentclass[doublespacing]{elsart}
\usepackage[english]{babel}
\usepackage{bm}

\usepackage{epsfig}
\usepackage{graphics}

\usepackage{amssymb}

\begin{document}

\begin{frontmatter}v


\title{A Benchmark Construction of Positron Crystal Undulator}

\author{V.V. Tikhomirov}
 \ead{vvtikh@mail.ru}
 \address{Research Institute for Nuclear Problems, Bobruiskaya str., 11, 220050, Minsk, Belarus}


\begin{abstract}
Optimization of a positron crystal undulator (CU) is addressed.
The ways to assure both the maximum intensity and minimum spectral
width of positron CU radiation are outlined. We claim that the
minimum CU spectrum width of 3 -- 4\% is reached at the positron
energies of a few GeV and that the optimal bending radius of
crystals planes in CU ranges from 3 to 5 critical bending radii
for channeled particles. Following suggested approach a benchmark
positron CU construction is devised and its functioning is
illustrated using the simulation method widely tested by
experimental data.
\end{abstract}



\end{frontmatter}
\section{Introduction}
Crystal undulators (CUs) \cite{bar,kap} open up wide possibilities
for designing sources of intense x- and $\gamma$-radiation. The
unique properties of CUs are assured by both the intra-atomic
strength and the inter-atomic space scale of the field of crystal
planes. After the modulation of transverse displacament of crystal
planes in CUs their field allows to induce the undulator-like
motion of channeled particles characterized by the record-breaking
acceleration and oscillation frequency. These unique properties
open up a way to generate intense hard narrow-spectrum radiation
using particle beams of reasonable energies.

Though CUs were proposed quite long ago \cite{bar,kap}, reliable
realistic description of their functioning has become possible
only after the development \cite{gui2,bar5} and experimental
validation \cite{ban,maz} of the corresponding simulation tool.

Both various approaches to CU fabrication and different views on
the previous achievements in CU studies have been addressed in
\cite{bar5,bel2,kor}. Our analysis \cite{bar5} of both conducted
and suggested experiments demonstrates limited perspectives of
electron CUs and stimulates more active investigation of the
positron case. In this paper we address the problem of finding an
optimal construction of positron CU suggesting the ways how to
increase the intensity and reduce the spectral width of the CU
radiation.

A method of optimal positron CU parameter search is outlined in
Section 2. Some details of our simulation method are discussed in
Section 3 along with the results of its applications to both the
recent test \cite{wis} of the short-period electron CU \cite{kos}
and 6.7 GeV channeling positron radiation experiment \cite{bak}.
Section 4 describes some benchmark positron CU configuration and
illustrates its functioning.


\section{CU parameter choice}\label{s2}

In order both to describe the necessary features of the positron
CUs and to devise a method to fix their parameters we will
consider the positron motion and radiation in the field of crystal
planes transversely modulated according to the function
\begin{equation}
\label{eq1} X(z) = A\cos \left( {\frac{2\pi }{\lambda _U }z}
\right) = A\cos (k_z z),\quad k_z = \frac{2\pi }{\lambda _U },
\end{equation}
where $A$ and  $\lambda_U$ are CU amplitude and period
respectively, x axis is normal and z axis is parallel to the
crystal planes before modulation.

The positron transverse coordinate $x(z) + X(z)$ measured in the
laboratory reference frame obeys relativistic Lorentz equation of
motion in the electric field of the modulated planes (\ref{eq1}).
This equation can be transformed to that for the positron
coordinate $x(z)$ measured from some modulated crystal plane and
governed by the effective planar potential $V_{eff}[x,R(z)]$ (see
Fig. 1), depending on the local crystal bending radius
\begin{equation}
\label{eq2} R(z) \approx \left( {\frac{d^2X(z)}{dz^2}} \right)^{ -
1} = - \frac{1}{\cos (k_z z)\;A\,k_z^2 },
\end{equation}
having the minimum absolute value
\begin{equation}
\label{eq3} R_{\min } = 1 / Ak_z^2.
\end{equation}

\begin{figure}
\label{Fig1}
 \begin{center}
\resizebox{110mm}{!}{\includegraphics{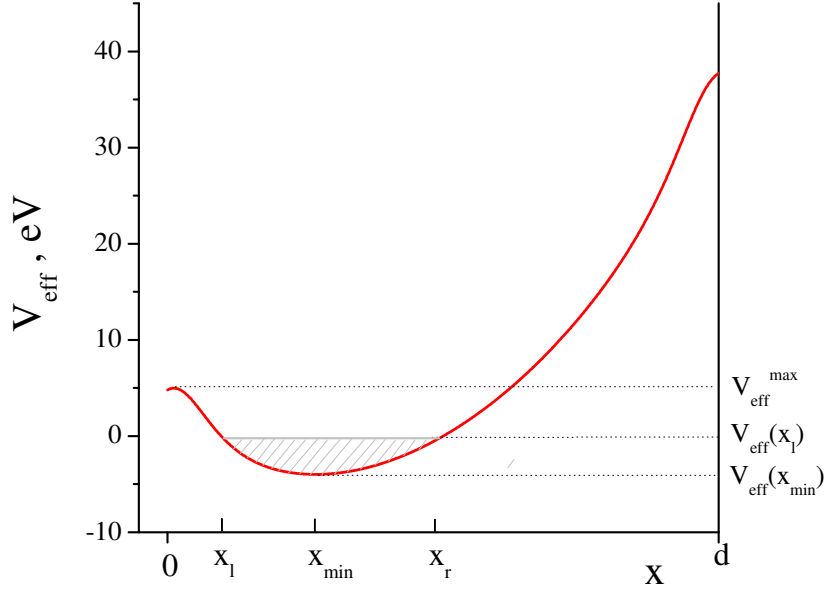}}\\
\caption{Dependence of the effective potential of bent crystal
planes on the distance $x$ from the atomic plane at the bending
radius R = 3.5 R$_{cr}$, x$_{r}$ and x$_{l }$ are the coordinates
of the right and left reflection points of the channeled positron,
x$_{min}$ is the coordinate of the local effective potential
minimum, $V_{eff}(x_{min})$ and $V_{eff}^{\max}$ are the local
minimal and maximum values of the effective potential inside the
given inter-planar interval and $V_{eff}(x_l)$ is the potential
value at the boundaries $x_l = x_d$ and $x_r$ of the hatched
region of stable channeling motion.} \vspace{1cm}
\end{center}
\end{figure}

In principle, since the bending radius (\ref{eq2}) depends on $z$,
the particle transverse motion in the effective potential is
accompanied by the transverse energy $\varepsilon_\bot =
\varepsilon v_x^2(x) / 2 + V_{eff}(x)$ variation, where
$\varepsilon$ and $v_x$ are the particle total energy and
transverse velocity, respectively. However, since the planar
potential is close to the harmonic one far from the planes,
$\varepsilon_\bot$ will not deviate considerably from its initial
value $\varepsilon_{\bot0} = \varepsilon v_{x0}^2 / 2 +
V_{eff}(x_0)$ for a stably channeled positron.  We will also rely
on the fact that $\varepsilon _\bot$ will periodically return to
the close values with the CU period $\lambda_U$ in any realistic
potential when $\lambda_U$ substantially exceeds the channeling
period $\lambda_{ch}$ and the adiabatic condition holds true.

Let us remind that channeling is possible in a bent crystal only
when its bending radius exceeds the critical value
\begin{equation}
\label{eq4} R_{cr} \approx \varepsilon / eE_{\max },
\end{equation}
where $E_{\max }$ is the maximal strength of planar electric
field, close to $6\cdot 10^9$ V/cm for Si(110). In general the
ratio
\begin{equation}
\label{eq5} r = \frac{R_{min}}{R_{cr}}  = \frac{E_{max}}{E_{eff}}
\end{equation}
determines both the depth and the width of the $V_{eff}$
channeling well (see Fig. 1), which, in their turn, limit both the
dechanneling length and channeling efficiency (percentage of
channeling particles). Below we will find out that the ratio
(\ref{eq5}) also determines the effective amplitude $E_{eff}$ of
the CU field.

To find the optimal value of the ratio (\ref{eq5}), we will
consider zero incidence case $\vartheta _{x0} = 0$ and assume that
stably channeling positrons should not approach the locations of
the nuclei closer than by a "dechanneling distance" $x_d \approx
0.02 nm$. For this the incidence point coordinate $x_0$ should
belong to the interval $\left[x_d, x_r \right]$, where $x_r$ is
the right reflection point of the channeling positron having the
left one $x_l = x_d$. At this the channeling capture probability
will be limited by the value $P_{ch} = (x_r - x_l)/d$ also
determined by the ratio (\ref{eq5}).

Perhaps the main difference of crystal undulators from the
"normal" magnetic one consists in the fast channeling oscillations
superimposed on the less frequent undulator ones. The amplitude of
the velocity oscillations $v_{x0}$ is related with the transverse
energy by the formulae\footnote{the $c=\hbar=1$ system of units is
used.}
\begin{equation}
\label{eq6} \varepsilon_{\bot} = \varepsilon v^2_{x0}/2.
\end{equation}
An initial value $\varepsilon_{\bot 0}$ of the latter is
determined by both the coordinate and angle of positron incidence,
while its further evolution is caused by the effective crystal
potential $V_{eff}$ oscillations induced by that of the local
crystal curvature as well as by the incoherent positron scattering
by crystal nuclei and electrons. The point is that in practise the
maximal value $\varepsilon^{max}_{\bot}$ of (6), measured from the
local $V_{eff}$ minimum, can not be reduced with the depth of the
potential well by the transition $R_{min} \rightarrow R_{cr}$
since both $P_{ch}$ and dechanneling length would nullify
therewith. On the other hand, the undulator parameter or a
normalized amplitude of the CU velocity oscillations
\begin{equation}
\label{eq7} K = Ak_z \gamma = \frac{eE_{eff} \lambda _U }{2\pi
\,m} = \frac{eE_{\max } }{2\pi m}\lambda _U \frac{R_{cr} }{R_{\min
} } \approx 0.031\lambda _U (\mu m)E_{eff}
\left(\frac{GV}{cm}\right)
\end{equation}
increases with $R_{min} \rightarrow R_{cr}$. Thus, the balance of
number of channeling positrons, proportional to their capture
probability $P_{ch}$ into the channeling regime, of the radiation
intensity, proportional to $K^2$ at small $K$, and of the
dechanneling length, proportional to $\varepsilon^{max}_{\bot}$,
has to determine the optimal ratio (\ref{eq5}) by maximizing the
product
\begin{equation}
\label{eq8} F_{opt}(r) = K^2 \varepsilon^{max}_{\bot} P_{ch}.
\end{equation}
The harmonic potential model readily allows one to find the
minimizing value $r^{harm}_{opt} = 2.5$, or $R_{min} = 2.5R_{cr}$
analytically. More reliable results of the numerical evaluation of
the optimization function (\ref{eq8}) in a realistic Moliere
potential are represented in Fig. 2 demonstrating that the optimal
value of the ratio (\ref{eq5}) increases from $r \simeq 3.5-4$ for
the smallest positron beam angular divergencies to $r > 5$ for the
larger ones. This circumstance will help us to choose the
benchmark CU configuration in Section 4.

\begin{figure}
\label{Fig2}
 \begin{center}
\resizebox{110mm}{!}{\includegraphics{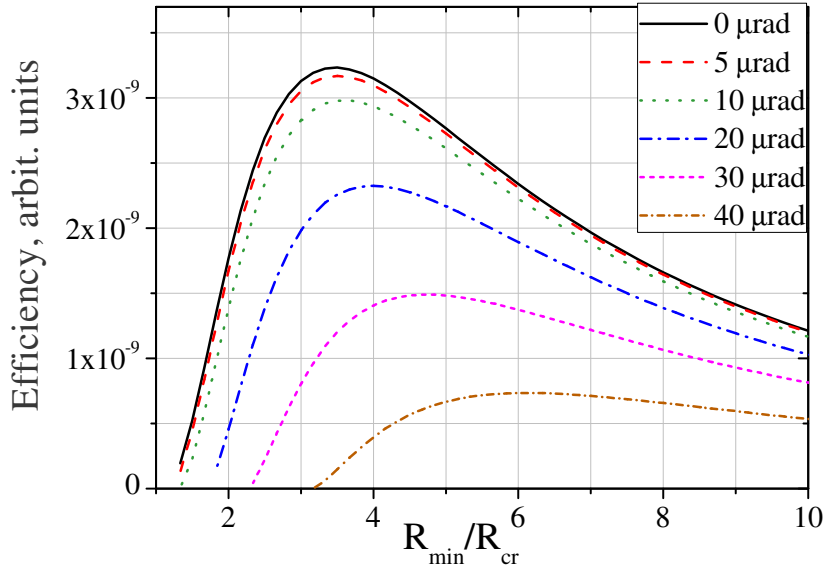}}\\
\caption{Optimization function (\ref{eq8}) dependence on the radii
ratio (\ref{eq5}) for indicated angles of positron incidence.}
\vspace{1cm}
\end{center}
\end{figure}

For an undulator, or any other periodic structure for coherent
radiation production, having a finite number $N$ of periods, the
radiation intensity (or probability) is proportional to the square
of the interference factor \cite{ell,nik,rul}
\begin{equation}
\label{eq9} F_N = \frac{\sin^2(N\Phi/2)}{\sin^2(\Phi/2)},
\end{equation}
where
\begin{equation}
\label{eq10} \Phi \approx \omega \lambda_U [\theta^2
+(1+K^2)/\gamma^2 + \varepsilon_\bot/\varepsilon]/2
\end{equation}
is a relative phase of the waves emitted from neighboring
undulator periods at an angle $\theta$ with respect to the CU axis
by a particle having a transverse energy $\varepsilon_\bot$.
Analysing the behavior of the interference factor in the usual way
\cite{ell,nik,rul}, one can represent the central emission line
frequency of the first harmonic of radiation in the forward
direction (at $\theta = 0$) in the form
\begin{equation}
\label{eq11} \omega_1 = \frac{\gamma^2}{1 + K^2/2 +
\varepsilon_{\bot 0} \, \varepsilon/m^2} \frac{2 \pi}{\lambda_U},
\end{equation}
where $\varepsilon_{\bot 0}$ is some effective average transverse
energy, determined by the $\varepsilon_{\bot}$ distribution both
in the channel and along the crystal length. The further analysis
of Eqs. (\ref{eq9})-(\ref{eq11}) allows one to obtain the
contributions by the limited number of CU periods
\begin{equation}
\label{eq12}  \frac{\delta\omega_N}{\omega_1} = \frac{1}{N}
\propto \frac{\lambda_U}{l_{dech}} \propto 1/\sqrt{\varepsilon},
\end{equation}
by the radiation collimation angle
\begin{equation}
\label{eq13} \frac{\delta\omega\,_\theta}{\omega_1} =
\frac{\theta^2 \gamma^2}{1 + K^2/2 + \varepsilon_{\bot \, 0} \,
\varepsilon/m^2}
\end{equation}
and by the transverse energy dispersion $\Delta\varepsilon_{\bot
max} = \varepsilon_{\bot max} - \varepsilon_{\bot min} =
\varepsilon_{\bot max}$
\begin{equation}
\label{eq14} \frac{\delta \omega\,_{\varepsilon_\bot}}{\omega_1} =
\frac{\Delta\varepsilon_{\bot \, max} \,  \varepsilon /m^2 }{1 +
K^2/2 + \varepsilon_{\bot \, 0} \,  \varepsilon /m^2} \propto
\varepsilon
\end{equation}
to the width of the main undulator radiation peak.

Before using Eqs. (\ref{eq12})-(\ref{eq14}) to specify further the
CU parameters, we initially fix the CU length to be close to that
of dechanneling and the CU period to range from three to four
channeling periods to assure, on the one hand, an adiabaticity of
the channeling motion in the undulator and, on the other, the
hardest radiation at the given positron energy. As a result one
finds that $\delta\omega_N/\omega_1 \propto 1/\sqrt{\varepsilon}$
(see Eq. (12)). Assuming that $K \sim 1$ and neglecting the last
term in the Eqs. (\ref{eq14}) denominator, one can see that
$\delta \omega\,_{\varepsilon_\bot}/\omega_1 \propto \varepsilon$
(see Eq. (14)). The energy dependencies of these estimates
indicate that a positron energy exists which minimizes the
combined radiation frequency uncertainty. Unfortunately, it is
hardly possible to find exactly this energy using only simple
analytical estimates (12) -- (14), which allow one to limit the
optimal positron energy by the relatively wide interval from 1.5
to 4 GeV. In fact, only a joint exhaustive search of the optimal
values of positron energy, CU period, modulation amplitude and
length can solve this problem. To provide a benchmark for this
search we will chose the lowest value $\varepsilon = 1.5 GeV$ from
the mentioned interval, the value which was not touched in the
literature for some time -- see \cite{kor}.

\section{Overview and some details of the simulation method}\label{s1}

In general the treatment of particle radiation in CUs is much more
complex than that in usual magnetic undulators. First of all, one
should take into consideration the dependence of radiation
formation process on the variation of particle longitudinal
velocity proceeding from the channeling oscillations, the
amplitude of which changes both from particle to particle and
along the trajectory of each particle. Also the radiation of
channeled, never channeled, dechanneled and rechanneled paricles,
as well as dechanneling and rechanneling processes themselves
should be properly accounted for. All these features can be taken
into consideration without their individual study if one uses a
reliable method of realistic particle trajectory simulations and
direct sampling of the accompanying radiation.

Our simulation method is widely tested using experimental data.
The reached level of agreement of its predictions with the latter
can be understood from the comparison of our predictions with the
data \cite{backe} obtained in experiments on electron radiation in
the $40 \mu m$ CU \cite{bar5}, on the radiation accompanying
electron multiple volume reflection from the multitude of the
planes of a single-piece crystal \cite{ban}, and on the electron
channeling observation in a thin bent crystal \cite{maz}. Since
the method has been already described in \cite{bar5, gui2}, we
will dwell here only on the incoherent scattering and photon
refraction treatment.

To remind the advantages of our approach, let us start from the
model \cite{tar} which was the first one described the incoherent
scattering of particles moving along classical trajectories.
Recently it was applied to simulate the crystal assisted
collimation \cite{tar2}. Remind that incoherent scattering process
is treated in \cite{tar,tar2} using the common formulae for the
r.m.s particle scattering angles on nuclei and electrons to which
the local nucleus and electron densities are substituted. In order
to make the relative variations of both averaged crystal field
strength and scatterer densities small on each simulation step,
the latter have to be chosen rather short. However being applied
to the short trajectory steps, the formulae for the r.m.s.
scattering angles on both nuclei and electrons can return only
small values, demonstrating their inability to realistically
describe the single particle scattering at the angles exceeding
the r.m.s scattering angle within small trajectory steps (the
"large" angles) considerably.

Such an approach is internally contradictive since, though the
large-angle scattering events are rare, they nevertheless give the
main contribution to the r.m.s. angle of scattering within a small
trajectory step. No wonder that single scattering proves to be
essential for an adequate joint treatment of coherent and
incoherent radiation and pair production \cite{ter,bar2,tik,art}
as well as for the quantitative description of the volume capture
into the channeling regime of both negatively \cite{tik,maz} and
positively \cite{bir} charged particles.

Incoherent scattering, in fact, was treated as a single scattering
process long ago by M.L. Ter-Mikaelian \cite{ter}. However this
approach, being polar to \cite{tar,tar2}, is too idealized as
well. Indeed, when a particle is moving nearly parallel to a plane
or axis, its successive collisions with nuclei and electrons
become much more frequent, especially in the regions of maximum
local nucleus and electron number densities inside atomic planes
and strings, tens of times exceeding their average concentrations
in a crystal. That is why we claim that the most adequate way
\cite{bar2,tik} both to describe analytically and to simulate
numerically the Coulomb scattering at the smallest angles, is to
treat it as a multiple scattering process as follows.

In order both to follow the CB theory of \cite{ter} and to ensure
a proper transition to the well fitted GEANT4 \cite{Jeant}
simulation results in the case of amorphous medium, we will adopt
the parametrization of the singly-charged particle scattering
cross-section
\begin{equation}
\label{eq15} \frac{d\sigma}{d\Omega} = \frac{ Z^2 \alpha^2}{p^2
\beta^2}\frac{1}{(\vartheta^2 + \vartheta_1^2)^2}
\end{equation}
from the Thomas-Fermi atom theory. Here $p$ ($\beta c$) is the
momentum (velocity) of the particle,
\begin{equation}
\label{eq16} \vartheta_1 = \frac{\hbar}{p\, a_{TF}}\left[ {{\kern
1pt} 1.13 + 3.76 (\alpha Z/\beta)^2} \right]^{1 / 2}
\end{equation}
is the typical scattering angle at which the nucleus field
screening by electrons becomes important and $$a_{TF} = (9 \pi
^2/128 Z)^{1 /3} (\hbar/m e^2) = 0.88534 \, a_0/Z^{1 /3},
$$
where $a_0$ is the Bohr screening radius, is the screening length
suggested by the Thomas-Fermi theory.

M.L. Ter-Mikaelian \cite{ter} also predicted some suppression of
the incoherent scattering by the correlations in particle
collisions with crystal atoms. This effect is explained by an
effective merging of small-angle correlated deflections
accompanying a sequence of peripheral particle collisions with
ordered atoms of a crystal plane or axis into a smooth trajectory
bending. Since the latter is described by the "continuum
potential", the corresponding small-angle particle scattering
contribution should be excluded from the incoherent scattering
process (see also \cite{lyu}). Taking into consideration this
incoherent scattering suppression by the Debye-Waller factor, the
mean square of the particle multiple scattering angle at a unit
length, comprised by the events of single scattering at the angles
$\vartheta \leq \vartheta_2$, can be represented in the form
\begin{equation}
\label{eq17}
\begin{array}{l}
 \left\langle {\vartheta _s^2 (z)} \right\rangle / dz = n_N \int_0^{\vartheta_2} \int_0^{2 \pi} \frac{d\sigma}{d\Omega}
 [1 - \exp(- p^2 \vartheta^2 u_1^2)] d \varphi \vartheta d
 \vartheta \\ =  4\pi
\frac{Z^2 \alpha^2}{\varepsilon^2} n_N
 \times \left\{ \ln (1 + a) +  [1 - \exp( - a \cdot b)]/(1 + a) \right. \\ \left. + (1 + b)\exp (b)\, [E_1
(b(1 + a)) - E_1 (b] \right\},
 \end{array}
\end{equation}
were $$E_1 (x) = \int\limits_x^\infty {e^{ - t}dt / t},\quad \quad
a = \vartheta _2^2 / \vartheta _1^2 ,\quad and \quad b = p^2
\vartheta^2 u_1^2.$$ The mean squared multiple scattering angle
(\ref{eq17}) should be used to sample a cumulative small-angle
scattering deflection in both transverse planes using the
corresponding 2D Gaussian distribution and azimuthal symmetry
assumption (see, however, \cite{lyu}). The value $\vartheta_2$ is
some boundary angle between the single and multiple scattering
regions, the choice of which is discussed below. The "large angle"
scattering by $\vartheta > \vartheta_2$, complementary to the
process of multiple scattering, is sampled as single scattering
events using the same cross-section (\ref{eq15}) at the instants
of time when a specially sampled random number exceeds the
single-scattering probability integrated along the trajectory from
either the previous single scattering event or the particle
entrance into the crystal. The suppression of incoherent
scattering by the atomic correlations in crystals is still taken
into consideration by discarding the scattering events in which
specially sampled random numbers do not exceed the value $\exp(-
p^2 \vartheta^2 u_1^2)$. The maximum single scattering angle
$\vartheta_{max}$ either can be taken equal to the maximum elastic
scattering angle by a nucleus or restricted by geometrical and
statistical considerations.

The disclosed approach evidently reduces to the model \cite{tar}
at $\vartheta_2 = \vartheta_{max}$ and to the model \cite{ter} at
$\vartheta_2 = 0$. Though the latter describes the situation more
adequately, it requires more extensive simulations. Our point is
that some freedom of the boundary angle $\vartheta_2$ choice
exists which allows one to facilitate the simulation procedure.
Indeed, the separation of the single-atom scattering angles on
"small" and "large", divided by the boundary value $\vartheta_2$,
closely reminds that of impact parameters or atom excitation
energies in the Bohr-Bethe-Bloch theory of ionization losses. The
boundary parameters of the latter enter the logarithms of the
intermediate complementary formulae in such a way that they drop
out from the resulting formula for ionization losses and,
therefore, can be roughly estimated from the qualitative
considerations. Similarly one can expect for the weak dependence
of the scattering simulation results on $\vartheta_2$.

To choose the latter, let us consider the particle incoherent
scattering in the process of its complete deflection by one atomic
string or axis, having in mind in particular the positron
deflection by a low-index Si plane in the GeV energy region
important for our study of the CU radiation. One can apply the
logarithmic approximation to estimate the averaged square of the
scattering angle
$$ \vartheta^2_s(pl) = 4 \pi \frac{Z^2 \alpha^2}{\varepsilon^2}
\ln (\vartheta_2/\vartheta_1) \int_{pl}n_N dz,
$$
where $\int_{pl}n_N dz$ is the integral of nuclei concentration
along a half of the positron channeling oscillation, and the
incoherent scattering probability $$P_{sc}(pl) =
\frac{\vartheta^2_s(pl)}{\vartheta_2^2 \ln
(\vartheta_2/\vartheta_1)}. $$ The latter demonstrates that
choosing $\vartheta_2 \sim \vartheta_s(pl)$ one can reduce the
probability of positron single Coulomb scattering along the half
of the channeling period down to $P_{sc}(pl) \leq 1$ preserving
the main features of the incoherent scattering process.

The disclosed method of particle trajectory simulation is
applicable to a vast number of processes accompanying high-energy
particle motion through crystals. Here we will apply it to the
simulation of the spectra of positron radiation in some CU
construction. As usual, we will proceed from the formulae
\begin{equation}
\label{eq18} \frac{d^2N}{d\omega d\Omega } = \frac{\alpha \omega
d\omega }{8\pi ^2{\varepsilon }'^2}\left[ {\omega ^2\left| A
\right|^2 / \gamma ^2 + \;\left( {\varepsilon ^2 + {\varepsilon
}'^2} \right)\;\vert \vec {B}\vert ^2} \right]
\end{equation}
\begin{equation}
\label{eq19} A = \int\limits_{ - \infty }^\infty {\exp \{i\varphi
(t)\}} dt, \quad \vec {B} = \int\limits_{ - \infty }^\infty
{\left( {\vec {v}_ \bot (t) - \vec {\theta }} \right)\exp
\{i\varphi (t)\}} dt,
\end{equation}
\begin{equation}
\label{eq20}
\begin{array}{l} \varphi (t) = \omega {\kern 1pt} t - \vec {k}{\kern
1pt} \vec {r} = \int\limits_0^t {\dot {\varphi }({t}')\,} d{t}' =
\frac{{\omega }'}{2}\int\limits_0^t {\left[ {\gamma ^{ - 2} +
\omega _p^2 / \omega ^2 + (\vec {v}_ \bot ({t}') - \vec {\theta
})^2} \right]\,} d{t}',\\ {\omega }' = \frac{\omega \varepsilon
}{{\varepsilon }'},
 \end{array}
\end{equation}
derived by the semiclassical operator method by Baier and Katkov
\cite{bai}. Note that a thorough description of the gamma-photon
refraction has been implemented in (\ref{eq18})-(\ref{eq20}) by
the introduction of the "plasma" refractive index
\begin{equation}
\label{eq21} n = k / \omega = 1 - \omega _{pl}^2 / 2 \omega^2.
\end{equation}
Remind that the photon refraction influence on both channeling and
CU radiation is known since the predictions of these effects
\cite{dub,bar,dub2}. Its importance for CU radiation became even
more clear recently. Indeed, the experiments \cite{backe,backe2}
on electron radiation in the $40 \mu m$ CU have demonstrated some
evidence of a peak at photon energies $20-60$ keV which are much
softer than that of the expected first CU peak. However, to judge,
whether such a feature in so soft spectral regio represents
itself some new resonant effect, one should first describe the
influence of transition radiation from the target surfaces taking
into consideration the difference in the thicknesses of the CU
crystal and the amorphous target used to measure the radiation
background.

The gamma refraction is also essential for another family of CU
constructions. Namely, a group of technologies \cite{bar5} using
periodic micro scratches, grooves \cite{bag} or thin strips
applied to the crystal surface for production of the periodic
surface strain can be readily applied to the fabrication of CUs
with periods $\lambda_U \sim 1 mm$. The point is that even at such
high positron energies as 15 GeV, the CU radiation peak frequency
(\ref{eq11}) practically coincides with the typical energy $\gamma
\omega_{pl} \simeq 1 MeV$ at which both the density effect and
transition radiation are most important.

Another improvement of the method \cite{bar5} concerns the
description of the contributions of the sharp incoherent and
smooth coherent deflections into the radiation process. Our recent
practice \cite{gui2,bar5} suggests to integrate (\ref{eq19}) by
parts over some trajectory intervals to separate the contributions
of sharp and smooth particle deflection. First -- as the
contributions of the integration interval ends and second --  as
the elementary integrals over the intervals analytically evaluated
using the uniform field approximation to assure a correct
high-frequency behavior without the interval length reduction
despite the coherent length decrease. Either the trajectory
simulation steps or the larger trajectory parts obtained by the
integration steps aggregation can be used as the trajectory
intervals. The disclosed approach allows one to represent the
integrals (\ref{eq19}) in the form
\[
\begin{array}{l}
 A = \int\limits_{ - \infty }^\infty {\exp \{i\varphi (t)\}} dt =
\frac{i}{\dot {\varphi }( + 0)} - \frac{i}{\dot {\varphi }( - 0)} + \\
 i\sum\limits_{i = 1}^N {\left\{ {\left[ {\frac{1}{\dot {\varphi }(t_i + 0)}
- \frac{1}{\dot {\varphi }(t_i - 0)}} \right]\exp i\varphi (t_i )
- \frac{2\ddot {\varphi }(\bar {t}_i )}{\dot {\varphi }^3(\bar
{t}_i )}\sin \left[ {\frac{\varphi (t_i - 0) - \varphi (t_{i - 1}
+ 0)}{2}} \right]\exp
i\varphi (\bar {t}_i )} \right\}} , \\
 \end{array}
\]
\begin{equation}
\label{eq22}
\begin{array}{l}
 \vec {B} = \int\limits_{ - \infty }^\infty {\left[ {\vec {v}_ \bot (t) -
\vec {\theta }} \right]\exp \{i\varphi (t)\}} dt = \left[
{\frac{i}{\dot {\varphi }( + 0)} - \frac{i}{\dot {\varphi }( -
0)}} \right]\left( {\vec
{v}_ \bot (0) - \vec {\theta }} \right)\; + \\
 i\sum\limits_{i = 1}^N {\left\{ {\begin{array}{l}
 \left[ {\frac{\vec {v}_ \bot (t_i ) + \vec {\vartheta }_i - \vec {\theta
}}{\dot {\varphi }(t_i + 0)} - \frac{\vec {v}_ \bot (t_i ) - \vec
{\theta
}}{\dot {\varphi }(t_i - 0)}} \right]\exp i\varphi (t_i ) - \\
 \frac{2}{\dot {\varphi }^2(\bar {t}_i )}\left[ {\dot {\vec {v}}_ \bot (\bar
{t}_i ) - \left( {\vec {v}_ \bot (\bar {t}_i ) - \vec {\theta }}
\right)\frac{\ddot {\varphi }(\bar {t}_i )}{\dot {\varphi }(\bar
{t}_i )}} \right]\sin \left[ {\frac{\varphi (t_i - 0) - \varphi
(t_{i - 1} + 0)}{2}}
\right]\,\exp i\varphi (\bar {t}_i ) \\
 \end{array}} \right\}} , \\
 \end{array}
\end{equation}
where ${\omega }' = \varepsilon / (\varepsilon - \omega )$, $\ddot
{\varphi }(t) = {\omega }'\left( {\vec {v}_ \bot (t_i ) - \vec
{\theta }} \right)\,\dot {\vec {v}}_ \bot (t)$, $\bar {t}_i = (t_i
+ t_{i - 1} ) / 2$, and the derivatives of the phase (\ref{eq20})
on the left and on the right of the entrance crystal surface
\[
\dot {\varphi }( - 0) = \frac{{\omega }'}{2}\left[ {\gamma ^{ - 2}
+ \left( {\vec {v}_ \bot (0) - \vec {\theta }} \right)^2} \right],
\]
\[
\dot {\varphi }( + 0) = \frac{{\omega }'}{2}\left[ {\gamma ^{ - 2}
+ \omega _p^2 / \omega ^2 + \left( {\vec {v}_ \bot (0) - \vec
{\theta }} \right)^2} \right];
\]
on the left and on the right of each inter-step border
\[
\dot {\varphi }(t_i - 0) = \frac{{\omega }'}{2}\left[ {\gamma ^{ -
2} + \omega _p^2 / \omega ^2 + \left( {\vec {v}_ \bot (t_i ) -
\vec {\theta }} \right)^2} \right],
\]

\[
\dot {\varphi }(t_i + 0) = \frac{{\omega }'}{2}\left[ {\gamma ^{ -
2} + \omega _p^2 / \omega ^2 + \left( {\vec {v}_ \bot (t_i ) +
\vec {\vartheta }_i - \vec {\theta }} \right)^2} \right]
\]
and on the right from the exit surface
\[
\dot {\varphi }(t_N + 0) = \frac{{\omega }'}{2}\left[ {\gamma ^{ -
2} + \left( {\vec {v}_ \bot (T) + \vec {\vartheta }_N - \vec
{\theta }} \right)^2} \right].
\]
have been used. We assume here that a particle intersects a plane
crystal from the left to the right. The introduced phase
derivative breaks at the crystal surfaces and at the trajectory
step boundaries allow to treat transition and incoherent
radiation, respectively.

Any available experiment, in that number the ones
\cite{backe,backe2,wis} on the electron CU radiation, should be
used both to verify the simulation tool and to demonstrate the
scope of its applicability. The outcomes of the experiment
\cite{backe} with 855 MeV electrons have been already reproduced
in \cite{bar5}. Both the modest radiation spectrum modification at
the expected CU peak position and the channeling radiation
suppression in the CU have been reproduced. Provided the
information on the real CU parameters, a quantitative reproduction
of this experiment will be readily given. The same is true for the
analogous experiments with 270 MeV \cite{backe} and 375 MeV
\cite{backe2} electrons in the same CU.

Here we reproduce the recent experimental results \cite{wis}
obtained with the short-period small-amplitude CU suggested by A.
Kostyuk \cite{kos}. From Fig. 3 one can conclude that the short
period small amplitude crystal plane modulation induces the
radiation spectrum modification near 15 MeV by about 10\% of the
height of the channeling radiation maximum at 5 MeV. More strict
reproduction of the experiment \cite{wis} is hampered by the
insufficient information on the actual parameters of both the
undulator and the electron angular distribution.
\begin{figure}
\label{Fig3}
\resizebox{110mm}{!}{\includegraphics{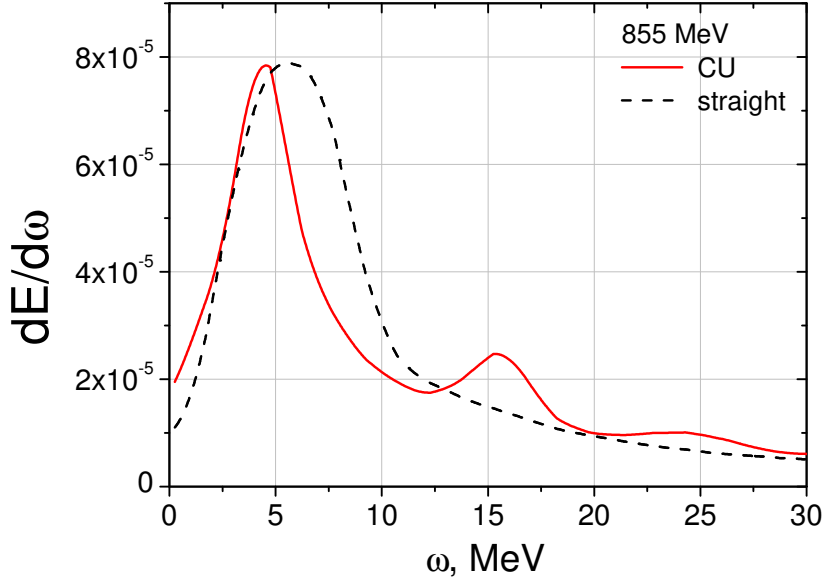}}\\
 \begin{center}
\caption{Spectral distribution of 855 MeV electron radiation in a
short-period CU with $\lambda_U = 4.3\mu m$, $A = 0.013 nm$ and
length of $3 \mu m$ (solid) and in a plane $Si (110)$ crystal of
the same length. Electron beam divergence equals $176 \mu rad$.}
\vspace{1cm}
\end{center}
\end{figure}

\begin{figure}
\label{Fig4}
 \begin{center}
\resizebox{110mm}{!}{\includegraphics{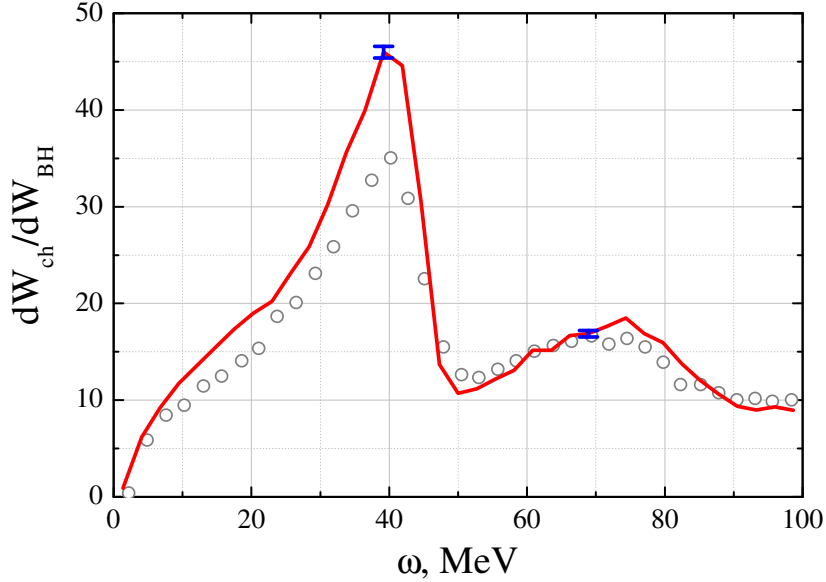}}\\
\caption{Enhancement factor over the Bethe–-Heitler radiation
intensity by that of the  6.7 GeV positrons channeling  in a 105
$\mu m$ thick Si (110) crystal. The incidence angles are
$-\theta_{ch} < \theta_{ex} < \theta_{ch},~ \theta_{ch} <
\theta_{ey} < \theta_{ch}$, where $\theta_{ch} = 62 \mu rad$. The
collimation angles are $-\theta_{coll} < \theta_{\gamma x} <
\theta_{coll},~ -\theta_{coll} < \theta_{\gamma y} <
\theta_{coll}$, where $\theta_{coll} = 350 \mu rad$. Open circles
represent experimental data from \cite{bak}. The bars represent
statistical simulation errors.} \vspace{1cm}
\end{center}
\end{figure}

The detail simulations of the electron channeling in both
\cite{bar5} and \cite{maz} reveales a drastic instability of the
electron channeling in CUs stimulating one to rely more on the
positron ones. Proceeding to the simulations of the latter, we
present the results of simulations of the positron channeling
experiment \cite{bak} in Fig. 4. Here again the reaching of the
better agreement of simulations with the experiment is hampered by
an incomplete information on the angular characteristics of both
the positron and gamma-photon beams.

\section{Simulation of a CU functioning}\label{s2}

\begin{figure}
\label{Fig5}
 \begin{center}
\resizebox{110mm}{!}{\includegraphics{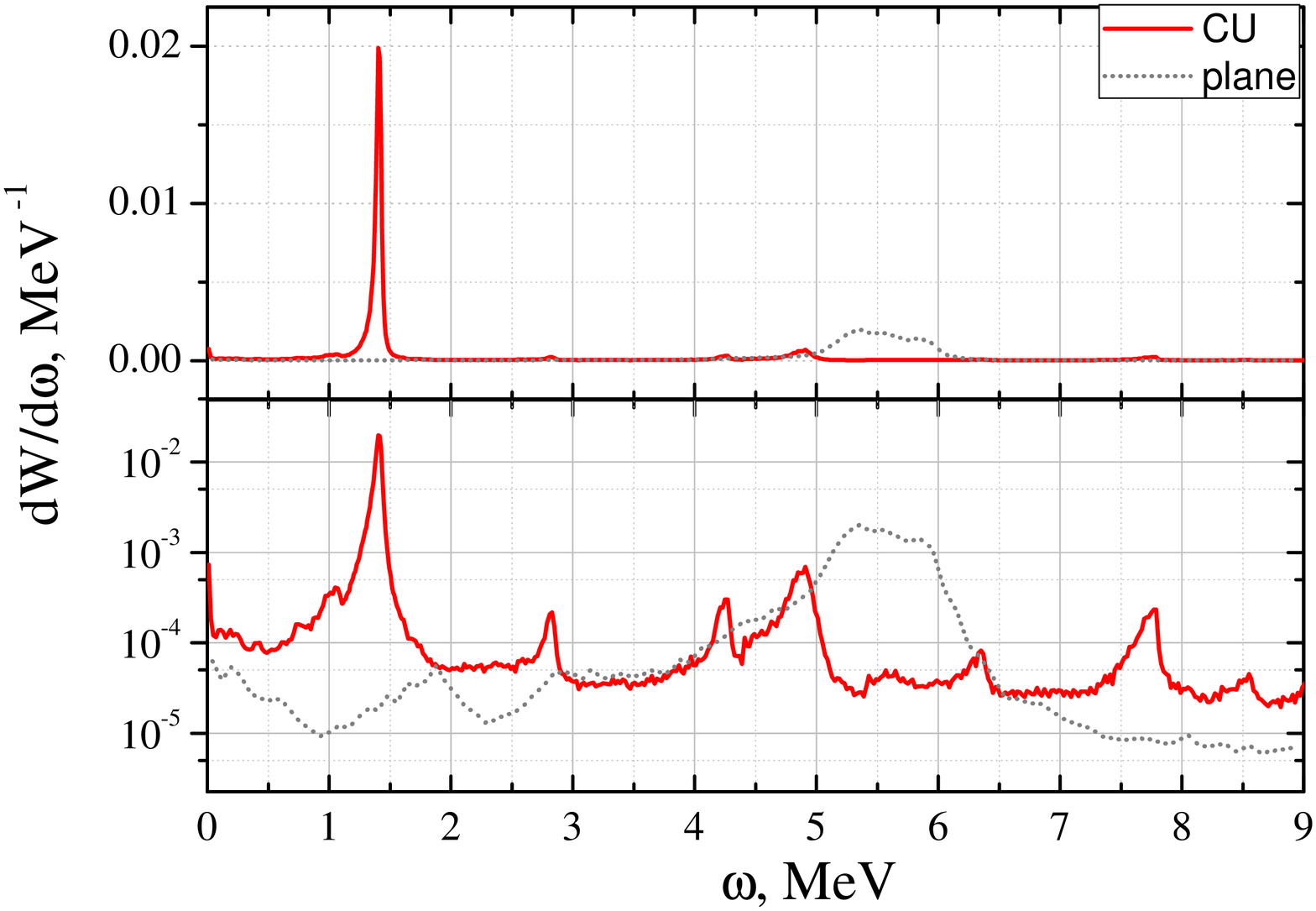}}\\
\caption{Spectral distribution of the radiation emitted by a 1.5
GeV positron in the linear (top) and logarithmic (bottom) scale:
solid line -- in the Si (110) CU (\ref{eq24}) and dotted line --
in a plane 0.48 mm Si (110) crystal. Positron beam incidence angle
equals zero, its angular divergence is $\Delta \theta_e = 10 \mu
rad$ and collimation semi-apex angle is $\Delta \theta_\gamma =
1/8\gamma = 42.6 \mu rad$.} \vspace{1cm}
\end{center}
\end{figure}

\begin{figure}
\label{Fig6}
 \begin{center}
\resizebox{110mm}{!}{\includegraphics{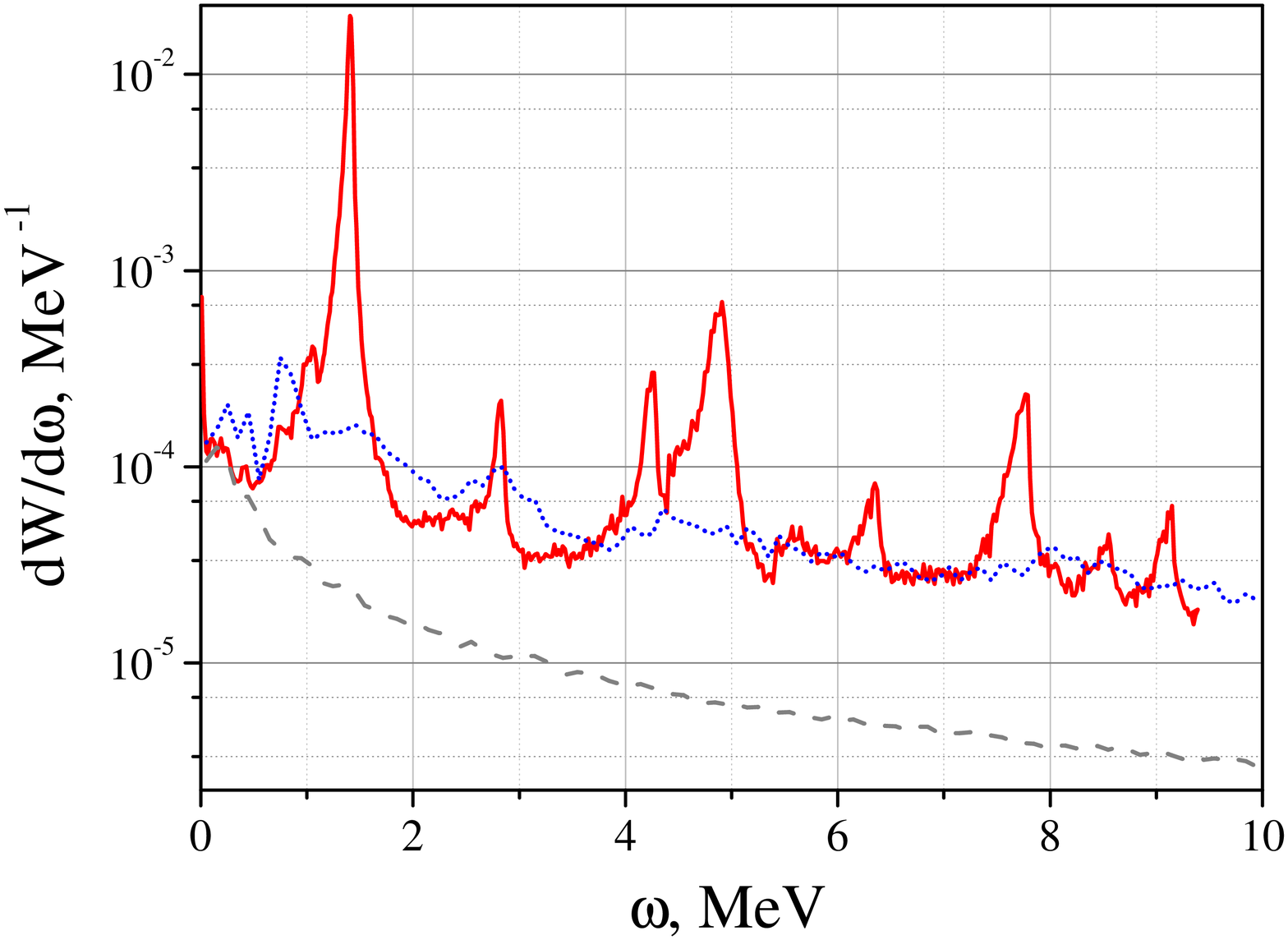}}\\
\caption{ Spectral distribution of the radiation: solid line --
emitted by the all positrons in the CU (24), dotted line -- by
only the initially nonchanneled positrons in the same CU and
dashed line -- by all the positrons in a 0.48 mm amorphous Si
plate. Positron energy equals 1.5 GeV. All the probabilities are
normalized on one positron.} \vspace{1cm}
\end{center}
\end{figure}

Some additional assumptions are necessary to apply eqs.
(\ref{eq5}), (\ref{eq7}) and (\ref{eq12})-(\ref{eq14}) to fix the
parameters of the benchmark CU assuring the lowest spectral width.
First we will choose the CU length to be slightly less than the
positron dechennling length. Also we will follow the traditional
understanding of the CU functioning, in which the positron
channeling in CU can be treated adiabatically, assuming $\lambda_U
= (3\div 4) \lambda_{ch}$. The choice of practically the minimal
$\lambda_U$ value, assuring adiabatical condition, provides both
the shortest undulator radiation wavelength, minimal undulator
length and positron energy as well as softens the requirements on
the positron beam angular divergence. Finding then a minimum of
either the sum of the widths Eqs. (\ref{eq12}) and (\ref{eq14}) or
of their squares and taking into consideration the numerous
uncertainties of all the assumptions made, one can both restrict
the "optimal" positron energy interval $\varepsilon = (1.5 \div 4)
GeV$ and fix the minimum radiation spectrum width
\begin{equation}
\label{eq23} \Delta\omega/\omega \simeq 3\%.
\end{equation}
Finding these estimates we have also assumed the $K \sim 1$ and $
R_{min} = 3.5 R_{cr}$. We chose the minimal positron energy from
the found interval which jointly with both the previous
assumptions and Eqs. (\ref{eq5}), (\ref{eq7}),
(\ref{eq12})-(\ref{eq14}) fixes the parameters
\begin{equation}
\label{eq24}
\varepsilon = 1.5 \, GeV,~~~ \lambda_U = 12 \, \mu m,~~~ A = 0.4
\, nm,~~~ L_U = 40 \lambda_U =0.48 \, mm
\end{equation}
of the CU scheme we would like to suggest here as a benchmark for
further investigations. It should be mentioned that $K = 0.64$ in
this construction. The results of the simulations of positron
motion and radiation in the CU (\ref{eq24}) are illustrated in
Figs. 5-8 which confirm the main prediction (\ref{eq23})
concerning the CU radiation spectrum width $\Delta\omega/\omega
\simeq 0.05/1.4 \simeq 3.5\%$ and enlighten some other some
important details.

\begin{figure}
\label{Fig7}
 \begin{center}
\resizebox{110mm}{!}{\includegraphics{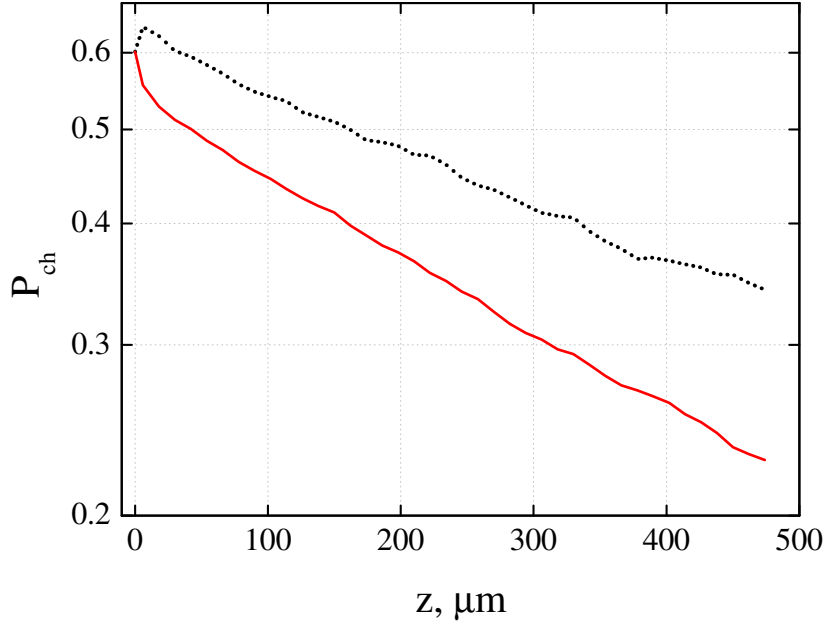}}\\
\caption{1.5 GeV positron channeling probability dependence on
crystal depth with (dotted, $l_{dch}$ = 0.8 mm) and without (solid
line, $l_{dch}$ = 0.56 mm) rechanneling consideration for the same
CU.} \vspace{1cm}
\end{center}
\end{figure}

This way Fig. 5 illustrates a great suppression of the channeling
radiation in the CU, primarily explained by both oscillation
amplitude reduction of the positrons stably channeling in CU and
moving them off the regions of the strongest planar field for a
considerable part of the CU period. Some frequency decrease of the
channeling radiation in the CU is readily explained by the joint
effect of the radiation collimation and variations of local plane
orientation in CU.

Fig. 6 demonstrates that the peak spectral intensity of radiation
of positrons channeling in the CU exceeds that of the comparable
number of nonchanneled ones by two orders of value and that in the
amorphous Si target of the same thickness -- by nearly three
orders of value.

Fig. 7 both confirms our assumption concerning the value of
positron dechanneling length and illustrates an importance of the
positron rechanneling, the effect recently observed for electrons
\cite{bar5,maz}.

\begin{figure}
\label{Fig8}
 \begin{center}
\resizebox{110mm}{!}{\includegraphics{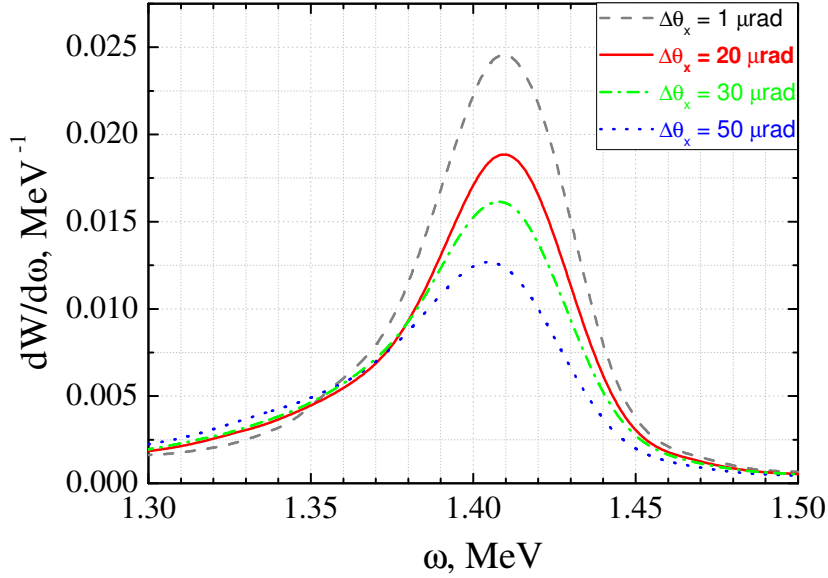}}\\
\caption{Spectral distribution of the radiation emitted in the
main peak region in the same CU at the indicated values of r.m.s
Gaussian positron beam divergences.} \vspace{1cm}
\end{center}
\end{figure}

Fig. 8 illustrates the role of incident beam divergence
demonstrating that the positron collimation requirements are quite
severe. The same is true for the emitted radiation. In order to
prevent a considerable spectral widening caused by the radiation
angular divergence we used the condition (\ref{eq13}) to restrict
the collimation angle to $\theta_{coll} = 1/8\gamma \simeq 43 \mu
rad$. Such a low value results in a rather low gamma-photon yield
of about $0.001\gamma/e^+$ which, however, can be increased, at
least several times at a prise of a moderate spectrum widening by
the two-three time increase of the collimation angle.

\section{Conclusions}

Relying on the original estimates of both the CU radiation
intensity and various contributions to the radiation spectrum
width a construction of positron CU, characterized by 3.5 -- 4\%
spectrum line width, $10^{-3}\gamma/e^+$ photon yield, an order of
value channeling radiation suppression and severe requirements on
both gamma and positron beam collimation is devised. We suggest to
consider this CU configuration as a benchmark for both estimates
of CU application perspectives and further development of
effective CU constructions.

The author is grateful to Prof. Baryshevsky and Dr. Maisheev for
useful discussions and to Prof. Guidi and his group for a
continuous collaboration.




\newpage

\begin{thebibliography}{199}


\bibitem{bar}
\noindent V.G. Baryshevsky, I.Ya.Dubovskaya, and A.O. Grubich,
Phys. Lett. \textbf{77A}(1980)61.

\bibitem{kap}
\noindent V.V. Kaplin, S.V. Plotnikov and S.A. Vorobijev, Zh.Tech.
Fiz. \textbf{50}(1980)1079 [English translation: Sov. Phys.-Tech
Phys. \textbf{25}(1980)650].

\bibitem{gui2}
V. Guidi, L. Bandiera and V.V. Tikhomirov, Phys. Rev. A
\textbf{86}(2012) 042903(11pp).

\bibitem{bar5}
V.G. Baryshevsky, V.V. Tikhomirov, Nucl. Instrum. And Methods. B
\textbf{309}(2013) 30.

\bibitem{ban}L. Bandiera et al., Phys. Rev. Lett. \textbf{111} (2013) 255502.

\bibitem{maz} A. Mazzolari  et al., Phys. Rev. Lett. \textbf{112} (2014) 135503.

\bibitem{bel2} S. Bellucci, V.A. Maisheev,  J. Phys.: Condens. Matter \textbf{18} (2006)
S2083–S2093.

\bibitem{kor}
 A.V. Korol,  A.V. Solov'yov and W. Greiner. Channeling and Radiation
in Periodically Bent Crystals (Springer Series on Atomic, Optical
and Plasma Physics 69).

\bibitem{wis}
T. N. Wistisen et al., Phys. Rev. Lett. \textbf{112} (2014)
254801.

\bibitem{kos}
A. Kostyuk, Phys. Rev. Lett. \textbf{110} (2013) 15503.

\bibitem{bak}
J. Bak et al., Nucl.Phys. \textbf{B254}(1985)491.

\bibitem{ell}
P. Elleaume, Beam Line 2002 \textbf{32} ¹ 1. P. 14.

\bibitem{nik}
 M.M. Nikitin, V. Ya. Epp, Undulator radiation. Moscow. Energoizdat,
1988. (in Russian).

\bibitem{rul}
R. Rullhusen, X. Artru, and P. Dhez, Novel radiation sources using
relativistic electrons. From infrared to x-rays. World Scientific
Publishing Co. Pte. Ltd., 1998.

\bibitem{backe}
H. Backe  et al., Nucl. Instrum. Methods Phys. Res.
\textbf{309}(2013)37.

\bibitem{backe2}
H. Backe  et al., J. Phys. Conf. Ser. \textbf{438}(2013)012017.

\bibitem{tar}
A.M. Taratin and S. A. Vorobiev, Sov. Phys. Tech. Phys.
\textbf{30}(1985)927.

\bibitem{tar2}
W. Scandale and A. Taratin, Simulation of "CRYSTAL", the bent
crystal based collimation experiment in the SPS, CERN/AT 2008-21.

\bibitem{ter}
M. L. Ter-Mikaelian, High-energy electromagnetic processes in
condensed media. Wiley. New York, 1972.

\bibitem{bar2}
V. G. Baryshevskii, V. V. Tikhomirov, Sov. Phys. JETP.
\textbf{63}(1986)1116.

\bibitem{tik2}
V. V. Tikhomirov, J. Physique. (Paris). \textbf{48}(1987)1009.

\bibitem{art}
X. Artru, Nucl. Instr. Meth. in Phys. Res. B \textbf{48}(1990)278.

\bibitem{tik}
V. V. Tikhomirov, Nucl. Instr. Meth. in Phys. Res. B
\textbf{36}(1989)282.

\bibitem{bir}
V.M.Biryukov, Y.A.Chesnokov, V.I.Kotov, Crystal Channeling and Its
Application at High-Energy Accelerators, Springer-Verlag, 1997,
219 p.

\bibitem{lyu}
V.L. Lyuboshitz, M.I. Podgoretskii, Sov. Phys. JETP.
\textbf{60}(1984)409

\bibitem{Jeant}
{http://geant4.cern.ch/} JEANT4 4 9.5.0 Physics Reference Manual
(6.70) - (6.73).

\bibitem{bai}
V. N. Baier, V.M. Katkov, and V. M. Strakhovenko, Electromagnetic
Processes at High Energies in Oriented Single Crystals (World
Scientific, Singapore, 1998).

\bibitem{dub}
V. G. Baryshevskii and  I. Ya. Dubovskaya, Phys. Status Solidi (b)
\textbf{82}(1977)403.

\bibitem{dub2}
V. G. Baryshevskii and  I. Ya. Dubovskaya, J. Phys.: Condensed
Matter \textbf{3}(14)(1991)2421.

\bibitem{bag}
E. Bagli et. al. Eur. Phys. J. C \textbf{74}(2014)3114.


\end{thebibliography}
\end{document}